\documentclass[useAMS,usenatbib]{mn2e}
\usepackage[utf8]{inputenc}
\usepackage{amsmath}
\usepackage{amssymb}
\usepackage{amsbsy}
\usepackage{graphicx}
\usepackage{units}
\usepackage{url}
\usepackage{color}

\newcommand{\bnabla}{\bmath{\nabla}}
\newcommand{\rbold}{\bmath{r}}
\newcommand{\Bbold}{\bmath{B}}
\newcommand{\vbold}{\bmath{v}}

\title[Protostellar collapse and fragmentation using an MHD GADGET]{Protostellar collapse and fragmentation using an MHD GADGET}

\author[F. Bürzle et al.]{Florian Bürzle$^{1}$\thanks{E-mail: florian.buerzle@uni-konstanz.de}, Paul C. Clark$^ {3}$, Federico Stasyszyn$^{2}$, Thomas Greif$^{2}$, \newauthor Klaus Dolag$^{2}$, Ralf S. Klessen$^{3,4}$ and Peter Nielaba$^{1}$\newauthor\\
$^{1}$Universität Konstanz, Fachbereich Physik, Universitätsstr. 10, 78464 Konstanz, Germany\\
$^{2}$Max Planck Institut für Astrophysik, Karl-Schwarzschild-Str. 1, 85741 Garching, Germany\\
$^{3}$Zentrum für Astronomie der Universität Heidelberg, Institut für Theoretische Astrophysik,\\ Albert-Ueberle-Str. 2, 69120 Heidelberg, Germany\\
$^{4}$Kavli Institute for Particle Astrophysics and Cosmology, Stanford University, Menlo Park, CA 94025, U.S.A} 
\begin{document}

\pagerange{\pageref{firstpage}--\pageref{lastpage}} \pubyear{2010}

\maketitle

\label{firstpage}
\bibliographystyle{mn2e}
\begin{abstract}
Although the influence of magnetic fields is regarded as vital in the star formation process, only a few magnetohydrodynamics (MHD) simulations have been performed on this subject within the smoothed particle hydrodynamics (SPH) method. This is largely due to the unsatisfactory treatment of non-vanishing divergence of the magnetic field. Recently smoothed particle magnetohydrodynamics (SPMHD) simulations based on Euler potentials have proven to be successful in treating MHD collapse and fragmentation problems, however these methods are known to have some intrinsical difficulties. We have performed SPMHD simulations based on a traditional approach evolving the magnetic field itself using the induction equation. To account for the numerical divergence, we have chosen an approach that subtracts the effects of numerical divergence from the force equation, and additionally we employ artificial magnetic dissipation as a regularization scheme. We apply this realization of SPMHD to a widely known setup, a variation of the 'Boss \& Bodenheimer standard isothermal test case', to study the impact of the magnetic fields on collapse and fragmentation. In our simulations, we concentrate on setups, where the initial magnetic field is parallel to the rotation axis. We examine different field strengths and compare our results to other findings reported in the literature. We are able to confirm specific results found elsewhere, namely the delayed onset of star formation for strong fields, accompanied by the tendency to form only single stars. We also find that the 'magnetic cushioning effect', where the magnetic field is wound up to form a 'cushion' between the binary, aids binary fragmentation in a case, where previously only formation of a single protostar was expected. 
\end{abstract}


\section{INTRODUCTION}
Magnetic fields, besides self-gravity, radiation and turbulence, are usually regarded as being the most fundamental constituents needed to describe star formation \citep[for recent reviews of theoretical aspects see, e. g.,][and references therein]{Mac-Low2004ys,McKee2007vn}. Especially the effects caused by magnetic fields, supported by increasing observational evidence of magnetic field structures in the interstellar medium (ISM) and molecular clouds \citep[see, e. g.,][and references therein]{Heilesuq}, came into the focus of interest within the past decade.

Numerical simulations, however, were not able to handle the full complexity of the physical processes connected with star formation for a long time, and were therefore limited to pure hydrodynamical, self-gravitating investigations. But in recent years the situation changed dramatically. Eulerian codes, which were always being able to handle magnetohydrodynamics (MHD) with good accuracy, became, with the advent of adaptive mesh refinement \citep[AMR, see][]{Berger1989fk}, able to handle protostellar collapse with MHD. Examples include the investigations by \cite{Ziegler2005fk} and \cite{Fromang2006kx}, who included collapse problems based on variations of the standard isothermal test case \citep{Boss1979fk}, to test their AMR codes, {\small NIRVANA} and {\small RAMSES} \citep{Teyssier2002ve}, respectively. They found a strong influence of magnetic fields on protostellar fragmentation in the limit of ideal MHD, that is, in a medium with infinite conductivity. Further investigations were performed by \cite{Commercon2010uq}, emphasizing the importance of considering the combined effects of MHD and radiative transfer on collapse and fragmentation. 

Also based on AMR, \cite{Machida2004ij} and \cite{Machida2005ab,Machida2005cd} performed several collapse simulations in ideal MHD, and found that fragmentation was suppressed by the magnetic field, but occurred still. More recently, these studies were extended to the formation of metal-free Population III stars in the early universe \citep{Machida2008zr}. 

In a series of publications, \cite{Hennebelle2008uq} and \cite{Hennebelle2008fk} also investigated the effect of magnetic fields on the collapse of dense molecular cloud cores. The former concentrates on magnetic braking and launching of outflows, where models with different magnetic field strengths are considered. For weak fields, they found negligible magnetic braking and thus formation of a centrifugally supported disc which in turn triggers a slowly expanding magnetic tower. For higher magnetic field strengths, they did not find formation of a centrifugally supported disc as a consequence of strong magnetic braking and collapse along the field lines. The latter publication however, focuses on fragmentation where a perturbation is added to the same setup as in the former work. Here, at weak field strengths, the centrifugally supported disc, which fragments in the hydrodynamic case, is found to remain stable and axisymmetric. For strong magnetic fields, again, no centrifugally supported disc is found because of magnetic braking and fragmentation is only found for strong initial perturbation amplitudes. 

Jets and outflows, however, are closely associated with star formation, and so many MHD simulations have been performed to investigate these phenomena, which are thought to be driven by coupling to magnetic fields. The study of jets and outflows is a field on its own right, so we refer to \cite{Banerjee2009ab}, and references therein, for a comprehensive discussion.

For more than 20 years, there has been an attempt to include MHD in smoothed particles hydrodynamics (SPH) for use on collapse problems, starting with the work of \cite{Phillips1986uq,Phillips1986fk}. However, his code lacks important algorithmic features developed afterwards and regarded as vital ingredients in SPH codes today, like adaptive smoothing lengths. More seriously, he considered only non-rotating clouds, so the results, which show no fragmentation either in the magnetized nor in the non-magnetized clouds, have to be taken with a grain of salt. 

\cite{Hosking2004kx} used a different approach based on a two-fluid formalism. This allowed non-ideal effects, such as ambipolar diffusion, to be taken into account, and thus enabled them to start from a subcritical rotating cloud core which, after following the evolution for some time, turned supercritical as result of diffusion. Their general conclusion was, that magnetic fields inhibit fragmentation. But their implementation suffered from non-zero divergence of the magnetic field, which did not allow them to follow the evolution for a long time.

An important step forward was the study by \cite{Price2007tg}. Their implementation is based on Euler potentials \citep{Stern1970fk,Rosswog2007kl}, which are free of physical divergence by construction. In their work, they considered two well known models, namely the axisymmetric collapse of a homogeneous density sphere and a variant of the \cite{Boss1979fk} 'standard isothermal test case' with initial $m=2$ perturbations in density.  They found, that stronger magnetic fields caused delays to the collapse, since the additional magnetic pressure provides additional support against gravity. Furthermore, they pointed out, that potentially crucial effects on discs might be caused from this delay, since the rate of mass infall onto the disc is reduced in this case. With respect to the perturbed clouds, these authors drew the main conclusions that magnetic fields might not be a serious problem to binary formation but that they suppress fragmentation. The latter is, contrary to previous results reported in the literature, attributed to the additional support by magnetic pressure, rather than magnetic tension forces or magnetic braking. 

In their following works, they turned to magnetic fields in cluster formation \citep{Price2008cr}. Using a barotropic equations of state, they found differences compared to pure hydrodynamical runs. Especially a significant influence to the star formation rate was reported by these authors. They performed further investigations by replacing the equation of state with a radiation transfer treatment, based on the flux-limited diffusion approximation \citep{Price2009ys}. As a main result of their investigations, the authors conclude that the net result of magnetic fields and radiative transfer is able to explain the inefficiency of star formation, with a star formation rate of 10 per cent per free-fall time, which is in good agreement with observations.  

However, despite their obvious success, it must be noted that Euler potentials have limitations on their own. The most serious one is that magnetic helicity is constrained to be zero, since the vector potential is always perpendicular to the magnetic field. In practice, this means that certain field configurations, namely such that are multi-valued as, e. g., combinations of poloidal and toroidal geometries, cannot be represented by Euler potentials. From this it follows, that such configurations also cannot be generated during a simulation, making it impossible to study certain physical processes, e. g., winding up of magnetic fields as found in dynamos or in protostellar outflow phenomena. Another limitation was pointed out recently by  \cite{Brandenburg2010fk}, who stressed the fact that Euler potentials are not able to deal with non-ideal MHD since even a small amount of diffusivity prevents convergence to the correct solution.

So other ways of dealing with the magnetic fields need to be considered, and the research on this topic is still ongoing. Another promising idea is an approach based on the vector potential, and indeed for spacial dimensions smaller than three, good results have been obtained by \cite{Price2010kx}. This is due to the fact that in these cases, the vector potential is in fact mathematically equivalent to a formulation with Euler potentials. However, \cite{Price2010kx} also showed that the vector potential formulation was not even able to handle standard test cases in three dimensions, causing him to suggest not to use this approach in an SPH context.

In this paper, we follow a different approach using the MHD implementation into the widely used {\small GADGET} code \citep{Springel2001il,Springel2005tw}, which has been applied successfully to several problems in galactic astrophysics \citep{Dolag2009uq,Kotarba2009fk,Kotarba2010ab}. We use the traditional approach for ideal MHD using the induction equation, but subtract the non-vanishing divergence term from the force equation to ensure numerical stability \citep{Borve2001zt}. As regularization scheme, we employ time-dependent artificial resistivity, introduced by \cite{Price2005ve}. We show, that this method produces accurate results which are not corrupted by non-vanishing divergence and compare well to some of the findings of \cite{Price2007tg}. However, at higher field strengths, we see noticeable deviations, the most prominent being the 'magnetic cushioning effect', where the magnetic field is wound up due to the rotation of the cloud and forms a cushion between the protostars.  The latter effect thus aids binary star formation in the case of a mass-to-flux ratio of $4$ (in critical units) where \cite{Price2007tg} had found just a single protostar.

\begin{figure}
\begin{center}
\includegraphics[width=9.5cm]{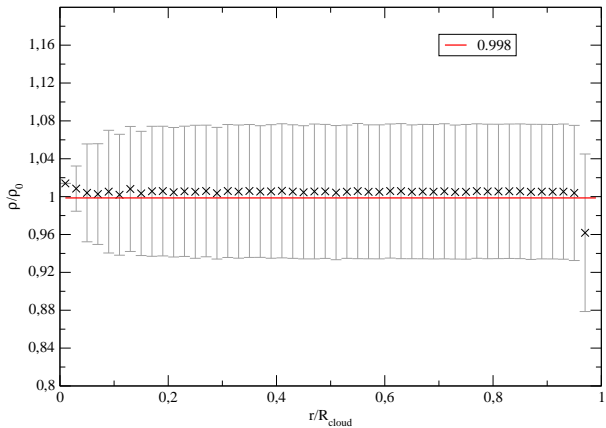}
\caption{\label{fig:scatter}The running average for the mean density at time $t=0$ is shown, where every cross indicates the mean within a bin of length $R/n_{\text{bin}}$ and the corresponding standard deviation is shown as error bar. We chose $n_{\text{bin}}=50$ for this analysis. The red line is a fit of the points to a constant function, obtaining a value of $0.998$ and thus a deviation from the analytical $\rho_0$ by only $0.2 \%$.}
\end{center}
\end{figure}

\begin{figure*}
\begin{center}
\includegraphics[width=\textwidth]{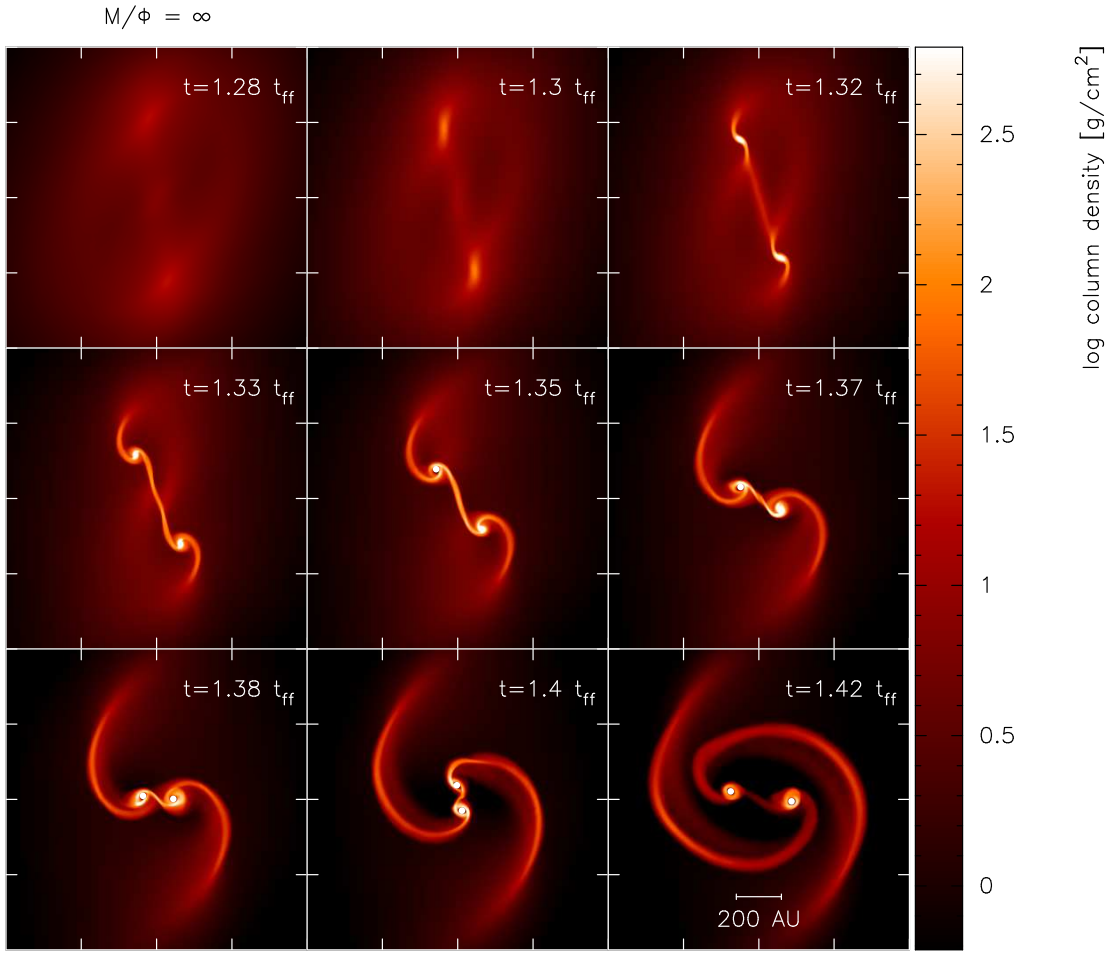}
\caption{\label{fig:hydro} Simulation results without magnetic field, i. e. the pure hydrodynamical case. In horizontal direction, $x$-coordinates are plotted, and $y$-coordinates are plotted vertically. Shown is the column density, here integrated along the $z$-axis, in physical coordinates. The panels display a sequence of increasing time, measured in in units of free-fall times, $t_{\text{ff}}=2.4\times 10^{4}\,\unit{yr}$.} 
\end{center}
\end{figure*}

\section{METHOD}
\subsection{Code}
For the (magneto-)hydrodynamical simulations presented in this work, we use the {\small GADGET} code \citep{Springel2001il,Springel2005tw}, a tree-based, massive parallel code utilizing the SPH method \citep[for recent reviews of SPH see, e.\ g.,][]{Rosswog2009qo,Springel2010ab}. The simulation results presented here were, as in \cite{Dolag2009uq}, obtained with the development version of {\small GADGET-3}. We used the code only in non-expanding, Newtonian space, so all equations referring to implementation details are lacking cosmological parameters and extensions.

In problems related to star formation, physical quantities vary over several orders of magnitude. Therefore, spatial and temporal adaptivity must be guaranteed within the simulation. The former is done using individual and adaptive smoothing lengths, where for each particle $i$ the equation

 \begin{equation}
\frac{4\pi}{3}h_i^3\rho_i = N m_i
\label{eqn:hsml}
\end{equation}
is solved iteratively with the density. Here, $h_i$ is the particles smoothing length, $N$ the number of neighbours, $m_i$ is the particle mass and $\rho_i$ is the density which is calculated according to

\begin{equation}
\rho_i = \sum_{j=1}^{N} m_j W(r_{ij},h_i)
\end{equation}
where $W$ is the cubic spline kernel \citep{MONAGHAN1985lc}. The dynamical equation

\begin{equation}
\left(\frac{\mathrm{ d}\vbold_i}{\mathrm{ d}t}\right)^{(\mathrm{hyd})} = - \sum_{j=1}^{N} m_j
\left[ f_i^\mathrm{ co
}\frac{P_i}{\rho_i^2}\bnabla_i W_i+f_j^\mathrm{ co
}\frac{P_j}{\rho_j^2}\bnabla_i W_j\right],
\end{equation}
has been derived using a variational principle \citep[e.\ g.][]{Springel2002uq}, and so the so called "grad $h$" correction terms 

\begin{equation}
f_i^\mathrm{ co}=\left[1+\frac{h_i}{3\rho_i}\frac{\partial
\rho_i}{\partial h_i }\right]^{-1},
\end{equation}
which account for the derivative of the kernel with respect to the smoothing length, are included by construction, ensuring energy and entropy conservation to time step accuracy.

To allow for accurate shock capturing, artificial viscosity is needed. The contribution of the viscous term to the particle acceleration is given by 

\begin{equation}
  \left(\frac{\mathrm{ d}\vbold_i}{\mathrm{ d}t}\right)^{(\mathrm{visc})} = -
\sum_{j=1}^{N} m_j\Pi_{ij}\bnabla_i\overline{W}_{ij},
\end{equation}

where $\overline{W}_{ij} = (W_i + W_j)/2$, and $\Pi_{ij}$ is the viscous tensor which is defined as 

\begin{equation}
\Pi_{ij}=
\begin{cases}
-\dfrac{\alpha v_{ij}^\mathrm{ sig}}{2\overline{\rho}_{ij}} \vbold_{ij}\cdot\hat{\rbold}_{ij} ,
 &\text{for } \vbold_{ij}\cdot\hat{\rbold}_{ij} \le 0,\\
0,&\text{for }\vbold_{ij}\cdot\hat{\rbold}_{ij} > 0.
\end{cases}
\end{equation} 
where $\overline{\rho}_{ij}=(\rho_i+\rho_j)/2$,  $\vbold_{ij} = \vbold_i - \vbold_j$ and $\rbold_{ij} = \rbold_i - \rbold_j$. This term was derived in close analogy to Riemann solvers \citep[see][]{Monaghan1997ff} and includes the signal velocity

\begin{equation}
v_{ij}^\mathrm{ sig} = c_i + c_j - \beta \vbold_{ij}\cdot\hat{\rbold}_{ij} ,
\label{eq:vsig}
\end{equation}
with the sound speed $c_i=\partial P_i/\partial \rho_i=\sqrt{\gamma P_i/\rho_i}$. For $\alpha$ and $\beta$ we use, as suggested by \cite{Dolag2009uq}, the values $2.0$ and $1.5$, respectively. Additionally, we would like to mention that we did not use the viscosity limiter introduced by \cite{Balsara1998ys}, since it is very likely responsible for introducing numerical artefacts in magnetic field growth, as observed in other work \citep{Kotarba2009fk,Kotarba2010ab}.

It should be noted, that artificial viscosity is a source of entropy (so is artificial resistivity, see below), which is generated at a rate $\text{d}A_i/\text{d}t$, where $A = P/\rho^{\gamma}$ is the entropic function. Since we use a barotropic equation of state, thus calculating the pressure directly as a function of density, an explicit consideration of the entropy production is not necessary and so we do not make use of the entropy treatment in {\small GADGET}. For a detailed discussion of the entropy formulation in SPH and implementation details, we refer to \cite{Springel2002uq} and \cite{Springel2005tw}.

The timestepping scheme is adaptive and an individual timestep for each SPH particle is chosen as a minimum of two criteria,

\begin{equation}
\Delta t_i = \min \left( \dfrac{2\eta\varepsilon_i}{|\mathbf{a}_i|},  \dfrac{C_\mathrm{ courant}h_i}{\mathrm{
max}_j(v_{ij}^\mathrm{ sig}) } \right).
\end{equation}
The first of these is based on a particles acceleration where $\eta$ is an accuracy parameter, with a numerical value of $10^{-3}$ in this work, and $\varepsilon$ the gravitational softening length, while the second is a Courant-like criterion that is needed to ensure numerical stability. For the sink particles (see section \ref{sinks}), only the first criterion is used.

\begin{figure*}
\begin{center}
\includegraphics[width=\textwidth]{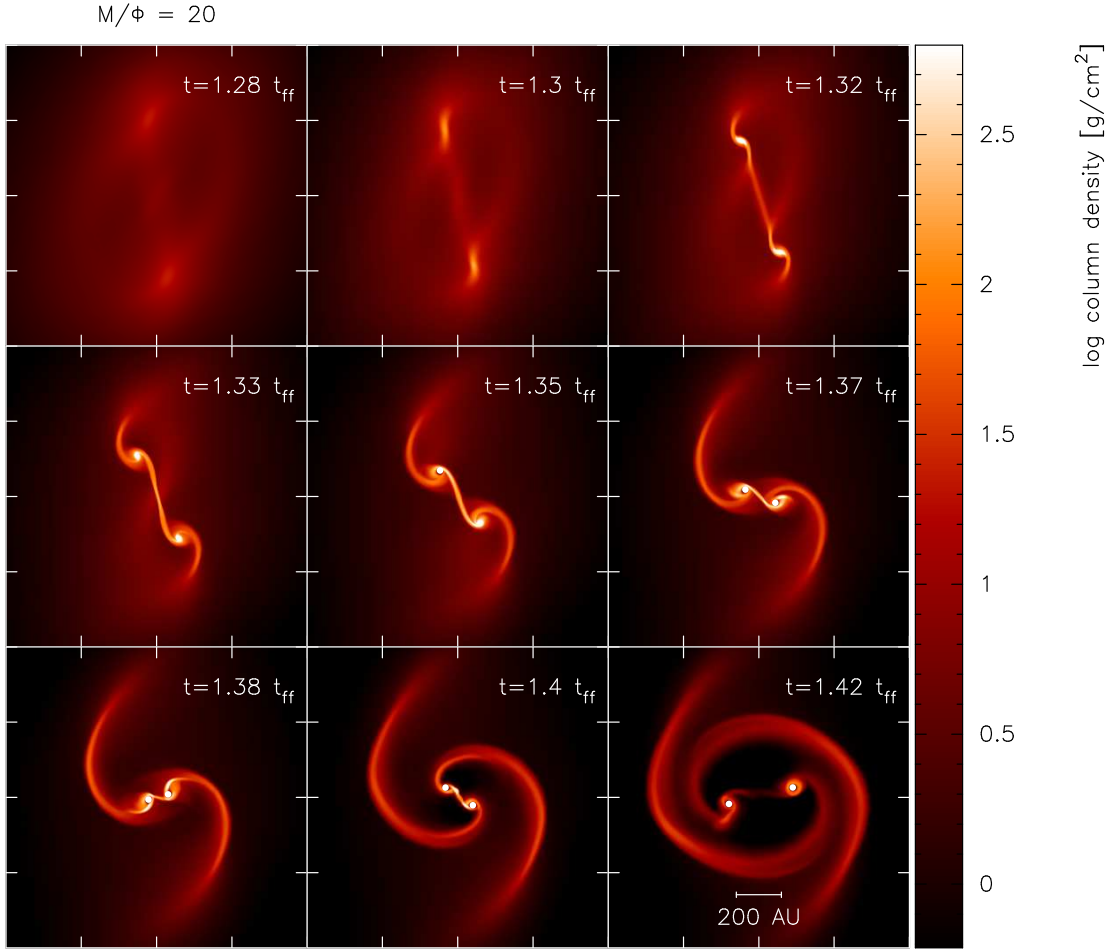}
\caption{\label{fig:mtof20} Simulation results with magnetic field parallel to the rotation ($z$) axis, with an initial field strength of $B_0 = 40.7\,\umu\unit{G}$. This corresponds to a mass-to-flux ratio of $M/\Phi = 20$, in multiples of the critical value. As previously, in horizontal direction, $x$-coordinates are plotted, and $y$-coordinates are plotted vertically. Shown is the column density, here integrated along the $z$-axis, in physical coordinates. The panels display a sequence of increasing time, measured in in units of free-fall times, $t_{\text{ff}}=2.4\times 10^{4}\,\unit{yr}$.} 
\end{center}
\end{figure*}

\subsection{Magnetohydrodynamics}

For the work presented here, we use the SPMHD implementation into {\small GADGET} described in detail in \citet{Dolag2009uq}. In the latter work, extensive tests on the reliability of the algorithms have been performed, among them some well known standard test cases typically used in the literature. These include the shocktubes considered in the work of \cite{Ryu1995zh}, the fast rotor by \cite{Balsara1999fk}, the Orszag-Tang vortex \citep{Orszag1979qz} and a variation of the strong blast test  \citep[e. g.][]{Balsara1999fk}. It was shown, that this implementation performs very well in general. Here we repeat the fundamental parts of the implementation and refer to \citet{Dolag2009uq} for a more detailed discussion on algorithms and performance in test cases.

For the correct capturing of shocks in the hydromagnetic case, it is essential to assign the correct artificial viscosity to the particles.  Therefore, the sound speed $c_i$ in eq.\ (\ref{eq:vsig}) is replaced by the speed of the fastest magneto-sonic wave 

\begin{equation}
v_i^{\text{(B)}} =
\frac{1}{\sqrt{2}}\left[\left(c_i^2+v_A^2\right)+
\sqrt{\left(c_i^2+v_A^2\right)^2 -
4\frac{c_i^2(\Bbold_i\cdot\hat{\rbold}_{ij})^2}{\mu_0\rho_i}}
\right]^{1/2}
\end{equation}
which enters also in the timestep criterion via the Courant condition.

The contribution of the magnetic field to the acceleration is given by

\begin{equation}
\left(\frac{\text{ d}v^k}{\text{ d}t}\right)^{(\mathrm{B})} = \dfrac{1}{\rho} \dfrac{\partial M^{kl}}{\partial x^l}
\label{eq:stress}
\end{equation}
where $M^{kl}$ is the magnetic stress tensor \citep{Phillips1985nq} defined as

\begin{equation}
M^{kl} = \dfrac{1}{\mu_0} \left( B^k B^l - \frac{1}{2}|\Bbold|^2\delta^{kl}\right)
\label{eq:tensor}
\end{equation}
where, as in equation (\ref{eq:stress}), the upper indices denote coordinates.
A straightforward discretization of the equation of motion, which also can be derived from a Langrangian using a variational principle \citep{Price2004dq}, is

\begin{equation}
\left(\frac{\mathrm{ d}\vbold_i}{\mathrm{ d}t}\right)^{(\mathrm{B})} = \frac{1}{\mu_0}
\sum_{j=1}^{N}m_j \left[f_i^\mathrm{ co}\frac{M_i}{\rho_i^2}
                       \cdot\bnabla_i W_i + f_j^\mathrm{co}\frac{M_j}{\rho_j^2}
                       \cdot\bnabla_j W_j \right]
\end{equation}
where $M_i$ and $\bnabla_i$, and the corresponding formulation for particle $j$, are abbreviations for $M_i^{kl}$ and $\bnabla_i^l$, respectively. While this formulation of the magnetic force conserves momentum exactly, it is also known to be unstable to negative stresses causing the particles to clump \citep{Phillips1985nq}. While many possible methods have been proposed in the literature to correct for this instability, most of them are rather impracticable or only of limited use, see \cite{Dolag2009uq} for a detailed discussion. In this work, we choose the formulation introduced by \cite{Borve2001zt} which subtracts the effect of any numerically non-vanishing divergence of the magnetic field. This is done by subtracting

\begin{equation}
\label{eq:borve}
\left(\frac{\text{ d}\vbold_i}{\text{ d}t}\right)^{(\text{corr})} =
\frac{\Bbold_i}{\mu_0}
\sum_{j=1}^{N}m_j \left[ f_i^\mathrm{ co}\frac{\Bbold_i}{\rho_i^2}
                       \cdot\bnabla_i W_i +
                       f_j^\mathrm{ co}\frac{\Bbold_j}{\rho_j^2}
                       \cdot\bnabla_j W_j \right]
\end{equation}
as in \cite{Dolag2009uq} where this method was used throughout and gave excellent results. Furthermore, the effects due to violation of momentum conservation have been shown to be negligible.

In ideal MHD, that is, in a medium with infinite electric conductivity, the magnetic field is advanced using the induction equation \citep{Price2004dq}

\begin{equation}
\dfrac{\text{d}\Bbold}{\text{d}t} = (\Bbold\cdot\bnabla)\vbold -\Bbold(\bnabla\cdot\vbold),
\end{equation}
and its SPH discretization is given by

\begin{equation}
\frac{\mathrm{ d}\Bbold_i}{\mathrm{ d}t} = 
\frac{f_i^\mathrm{ co}}{\rho_i}  
\sum_{j=1}^{N}m_j\left[
\Bbold_i(\vbold_{ij}\cdot\bnabla_i W_i)
-\vbold_{ij}(\Bbold_i\cdot\bnabla_i W_i)
\right].
\end{equation}
where the "grad $h$" correction terms are included for consistency (but note that this can not be derived from first principles).

Since a further important source of errors is the noise introduced by numerical fluctuations of the magnetic field, which originate in integration errors, a regularization procedure is required in the numerical scheme. We use artificial magnetic dissipation to regularize the underlying magnetic field. This is done in close analogy to the artificial viscosity by introducing a parameter $\alpha_B$ that controls the strength of the dissipative effect. As in \cite{Dolag2009uq} and \cite{Price2005ve}, the dissipative term is included into the induction equation

\begin{equation}
\left( \frac{\mathrm{ d}\Bbold_i}{\mathrm{ d}t} \right)^{\text{(diss)}}= 
\frac{\rho_i \alpha_B}{2}  
\sum_{j=1}^{N}
\dfrac{m_j v_{ij}^{\text{sig}}}{\overline{\rho}_{ij}} (\Bbold_i - \Bbold_j) \hat{\rbold}_{ij} \cdot \bnabla_i \overline{W}_{ij}.
\end{equation}
Note that $\alpha_B$ can be a constant or a time dependent quantity. In the latter case, a decay equation

\begin{equation}
\dfrac{\text{d}\alpha_B}{\text{d}t} = - \dfrac{\alpha_B - \alpha_B^{\min}}{\tau} + S
\end{equation}
is evolved for each particle, where the source term $S$ is given by

\begin{equation}
S = \dfrac{S_0}{\sqrt{\mu_0\rho}} \max \left( \left| \bnabla \cdot \Bbold \right| ,  \left| \bnabla \times \Bbold \right| \right).
\end{equation}
A natural choice for the characteristic decay time-scale $\tau$ is provided by considering the time a shock needs to travel through one kernel length and can therefore be written as \citep{Price2004qa}

\begin{equation}
\tau = \dfrac{h_i}{C \max \left( v_{ij}^\text{sig} \right) }.
\end{equation}
where $C$ is typically chosen in the same range as for the Courant timestep condition. In this work, we used the time-dependent version throughout, enforcing a maximum value of $\alpha_B^{\max} = 1.0$ for the resistivity parameter. The value of the magnetic field constant $\mu_0$ was chosen such, that the magnitudes of the magnetic fields are given in Gauss.

\begin{figure*}
\begin{center}
\includegraphics[width=\textwidth]{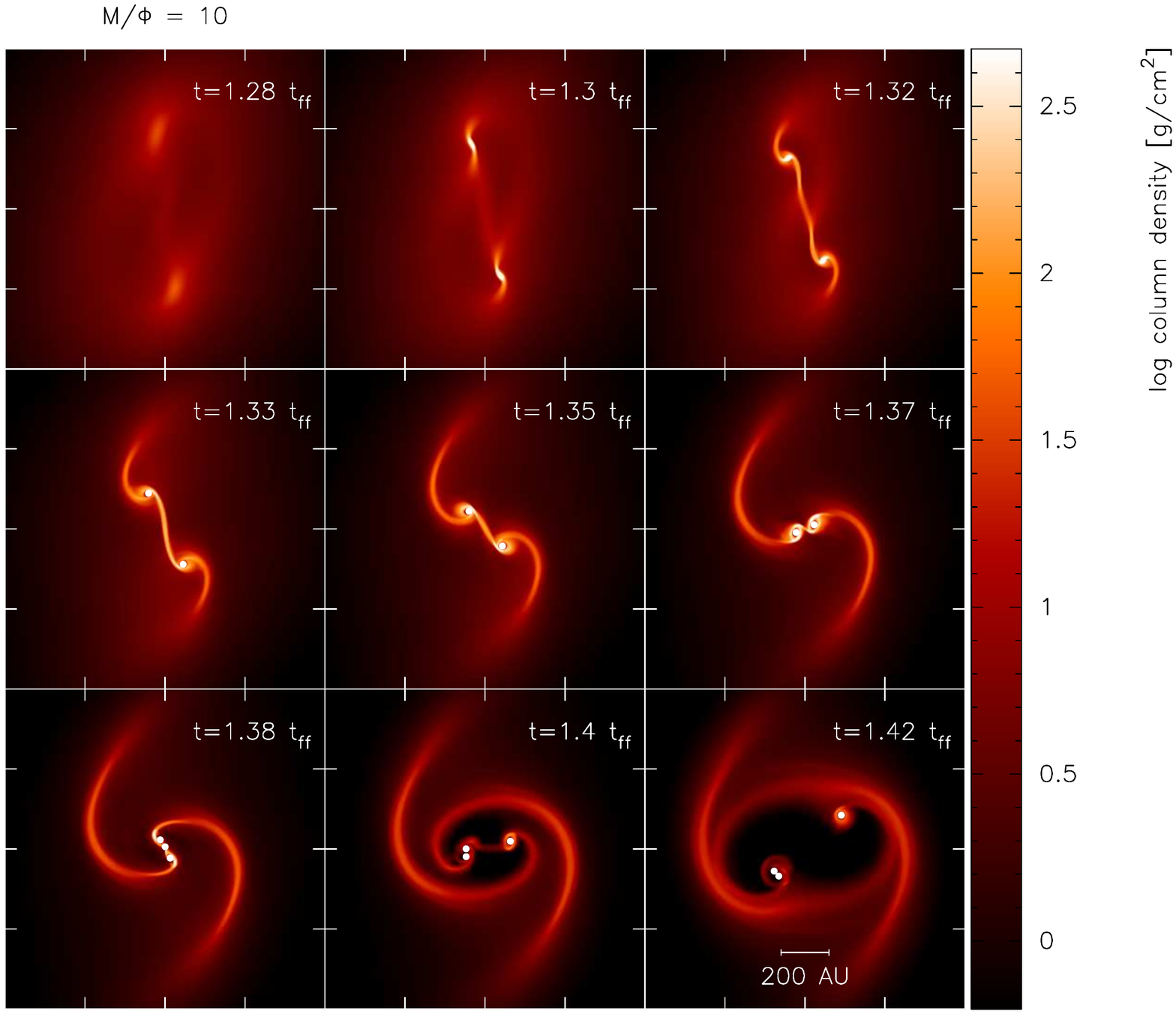}
\caption{\label{fig:mtof10} Simulation results with magnetic field parallel to the rotation ($z$) axis, with an initial field strength of $B_0 = 81.3\,\umu\unit{G}$. This corresponds to a mass-to-flux ratio of $M/\Phi = 10$, in multiples of the critical value. As previously, in horizontal direction, $x$-coordinates are plotted, and $y$-coordinates are plotted vertically. Shown is the column density, here integrated along the $z$-axis, in physical coordinates. The panels display a sequence of increasing time, measured in in units of free-fall times, $t_{\text{ff}}=2.4\times 10^{4}\,\unit{yr}$.} 
\end{center}
\end{figure*}

\subsection{Comments on artificial viscosity and artificial resistivity}
While artificial viscosity is needed to allow for correct shock capturing and to avoid unphysical effects such as particle interpenetrations, this approach certainly has its well known weaknesses. One of the most prominent, pointed out (among others) by \cite{Agertz2007ss}, is the fact that with viscosity present in the system, one effectively needs to solve the Navier-Stokes and not the Euler equations. To avoid errors resulting from this, one method is the utilization of time dependent artificial viscosity \citep{Morris1997lh} combined with a switch, so that viscosity can be reduced to a minimum if no sources of viscosity are present. However, this approach is problematic when applied to collapse problems, since the usual switch is based on the condition $\bnabla \cdot \vbold < 0$, indicating the presence of a shock. In context of a self-gravitating collapse, this condition is also a sign for a convergent flow and thus the switch might erroneously respond in this case. So since we assume here, that the errors due to the latter effect would be more serious in our application than errors from not solving the correct Euler equations, we decided to use a constant artificial viscosity throughout. However, to reduce the effect of intrinsical numerical diffusion, we followed the conclusions drawn by the work of \cite{Attwood2007rq}, and restricted the allowed range of the nearest neighbours $N$ to a value smaller than one. In this work, we used  $N = 64 \pm 0.3$ throughout.

The situation is similar with artificial resistivity. As already mentioned, artificial magnetic dissipation is needed to deal with noise related to magnetic fields. However, this means that in principle we need to solve the equations of non-ideal MHD to account for the additional diffusivity added by artificial resistivity, so using the time-dependent formulation is likely to improve the situation considerably. Furthermore, a constant dissipation of, say,  $\alpha_B = 1.0$, which is needed in later stages of the collapse would introduce a large scale initial diffusion especially in the low density region of our system and thus lead to a significant disturbance at early stages.

\begin{figure*}
\begin{center}
\includegraphics[width=\textwidth]{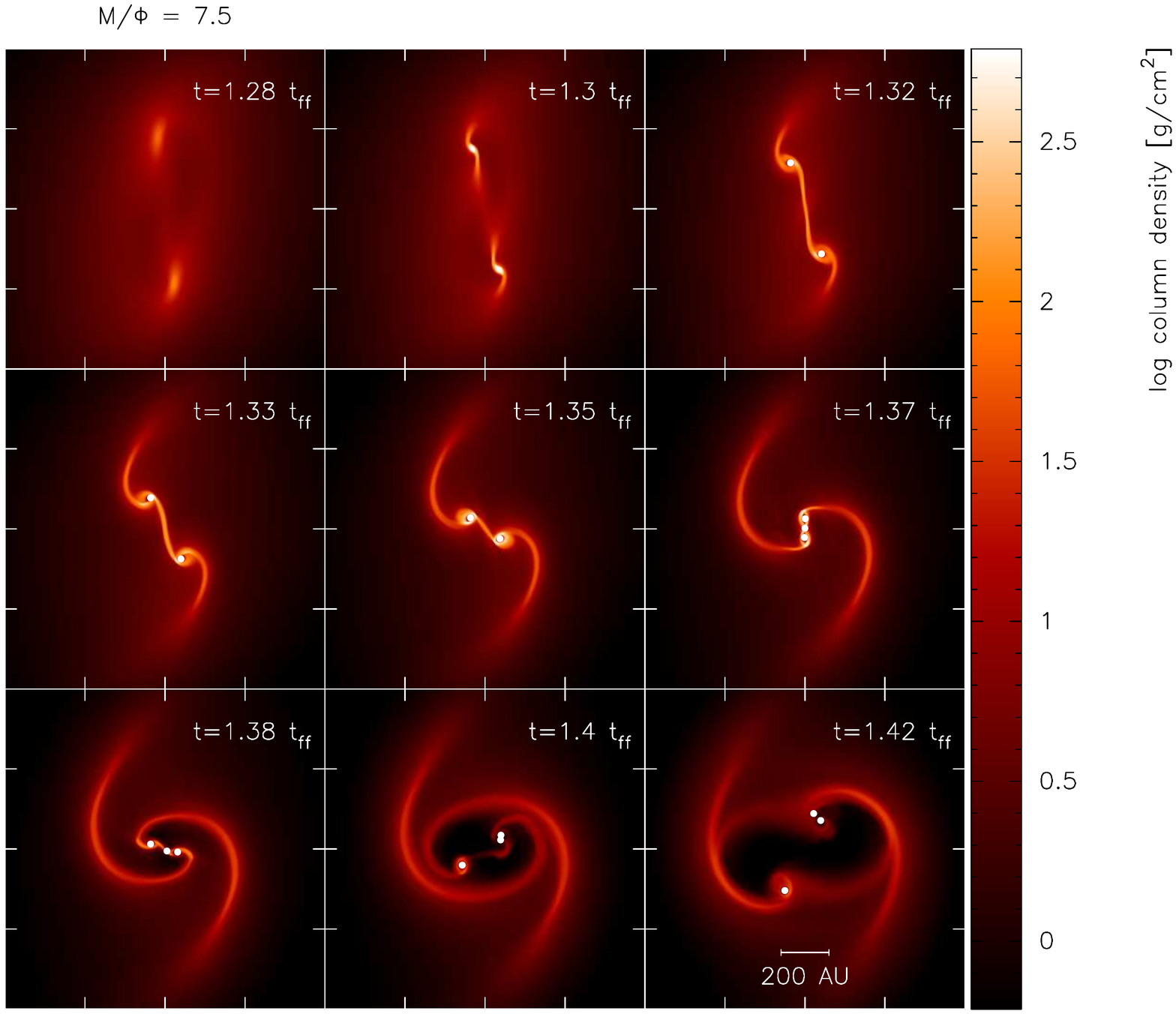}
\caption{\label{fig:mtof7.5} Simulation results with magnetic field parallel to the rotation ($z$) axis, with an initial field strength of $B_0 = 108.5\,\umu\unit{G}$. This corresponds to a mass-to-flux ratio of $M/\Phi = 7.5$, in multiples of the critical value. As previously, in horizontal direction, $x$-coordinates are plotted, and $y$-coordinates are plotted vertically. Shown is the column density, here integrated along the $z$-axis, in physical coordinates. The panels display a sequence of increasing time, measured in in units of free-fall times, $t_{\text{ff}}=2.4\times 10^{4}\,\unit{yr}$.} 
\end{center}
\end{figure*}

\subsection{Thermodynamics}
In our models, we use a piecewise equation of state, defined by

\begin{equation}
P=K\rho^{\gamma}.
\end{equation}
Since this equation of state is barotropic, i. e. the pressure is a function of the density only, the energy equation needs not to be solved explicitly. In this work, the adiabatic index is given by

\begin{equation}
\gamma=
\begin{cases}
1 ,
 &\text{for } \rho \le \rho_{\text{crit}},\\
7/5,
&\text{for }  \rho > \rho_{\text{crit}}.
\end{cases}
\end{equation} 
with $\rho_{\text{crit}} = 10^{-14} \unit{g}\, \unit{cm}^{-3}$. So for low densities the equation of state is isothermal and $K = c_{\text{s}}^2$, while for high densities it is adiabatic assuming a diatomic gas with five degrees of freedom. In the latter case, $K$ is chosen such that the pressure is continuous at the critical density, i. e. $K = c_{\text{s}}^2 \rho_{\text{crit}}^{-2/5}$.

\subsection{Sink particles\label{sinks}}
In those high density regions that are going to form a protostar, particles also gain large accelerations which in turn leads to assignment of very small time steps to a small fraction of the particles present in the whole system. Therefore, the timestep is becoming prohibitively small, and effectively causes the simulation to stall.
Sink particles, first introduced by \cite{Bate1995jh}, provide a way of solving this problem. When a certain threshold density is reached within some small region of space, characterized by the sink radius, the gas particles within this region are replaced by a non-gaseous particle that carries their masses and momenta. This particle interacts with other particles via gravity only, and is able to accrete further particles that cross its outer boundary. However, while making further evolution of  the collapse accessible, the method comes with the burden that all information within the sink particle is lost. But after all, sink particles have proven to be a very useful subgrid model which has been successfully applied in many star formation related studies. 

Since the pioneering work of \cite{Krumholz2004fk}, also Eulerian codes can benefit from sink particles, and recently \cite{Federrath2010fk} have accomplished a implementation into the widely used {\small FLASH} grid code \citep{Fryxell2000uq}. While the first implementation of sink particles into {\small GADGET-2} was done by \cite{Jappsen2005pz}, the {\small FLASH} implementation served as a prototype for our current implementation of sink particles in {\small GADGET-3}. 

In the studies presented here, we insert a sink particle, once a threshold density of $\rho_s=10^{-10} \unit{g}\, \unit{cm}^{-3}$ within an accretion radius of $\sim 13\, \unit{au}$ has been reached. We chose the creation density to be far in the adiabatic regime, to ensure that sinks are only formed in regions where the collapse has advanced already several orders of magnitude thus ruling out artefacts by spurious sink formation. The gas that is approaching a sink particle later, is accreted if it is bound to it and further criteria are met, as described in detail in \cite{Federrath2010fk}.

The treatment of the magnetic field in sink particles, however, has the same limitations as pointed out in \cite{Price2007tg,Price2008cr}, namely that magnetic field carried by accreted particles is discarded, so sink particles can not provide magnetic field driven feed back on the surrounding cloud.

\begin{figure*}
\begin{center}
\includegraphics[width=\textwidth]{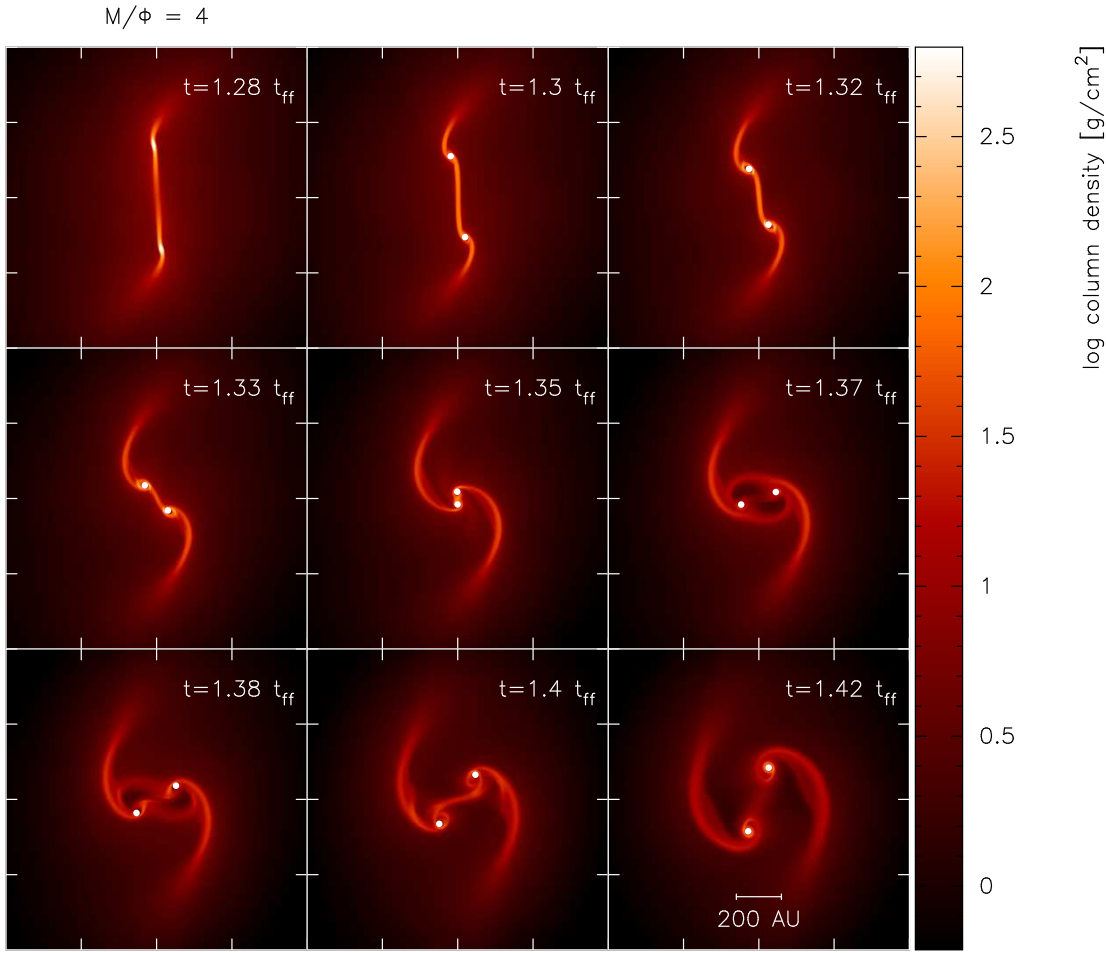}
\caption{\label{fig:mtof4} Simulation results with magnetic field parallel to the rotation ($z$) axis, with an initial field strength of $B_0 = 203\,\umu\unit{G}$. This corresponds to a mass-to-flux ratio of $M/\Phi = 4$, in multiples of the critical value. As previously, in horizontal direction, $x$-coordinates are plotted, and $y$-coordinates are plotted vertically. Shown is the column density, here integrated along the $z$-axis, in physical coordinates. The panels display a sequence of increasing time, measured in in units of free-fall times, $t_{\text{ff}}=2.4\times 10^{4}\,\unit{yr}$.} 
\end{center}
\end{figure*}

\begin{figure*}
\begin{center}
\includegraphics[angle=-90,width=\textwidth]{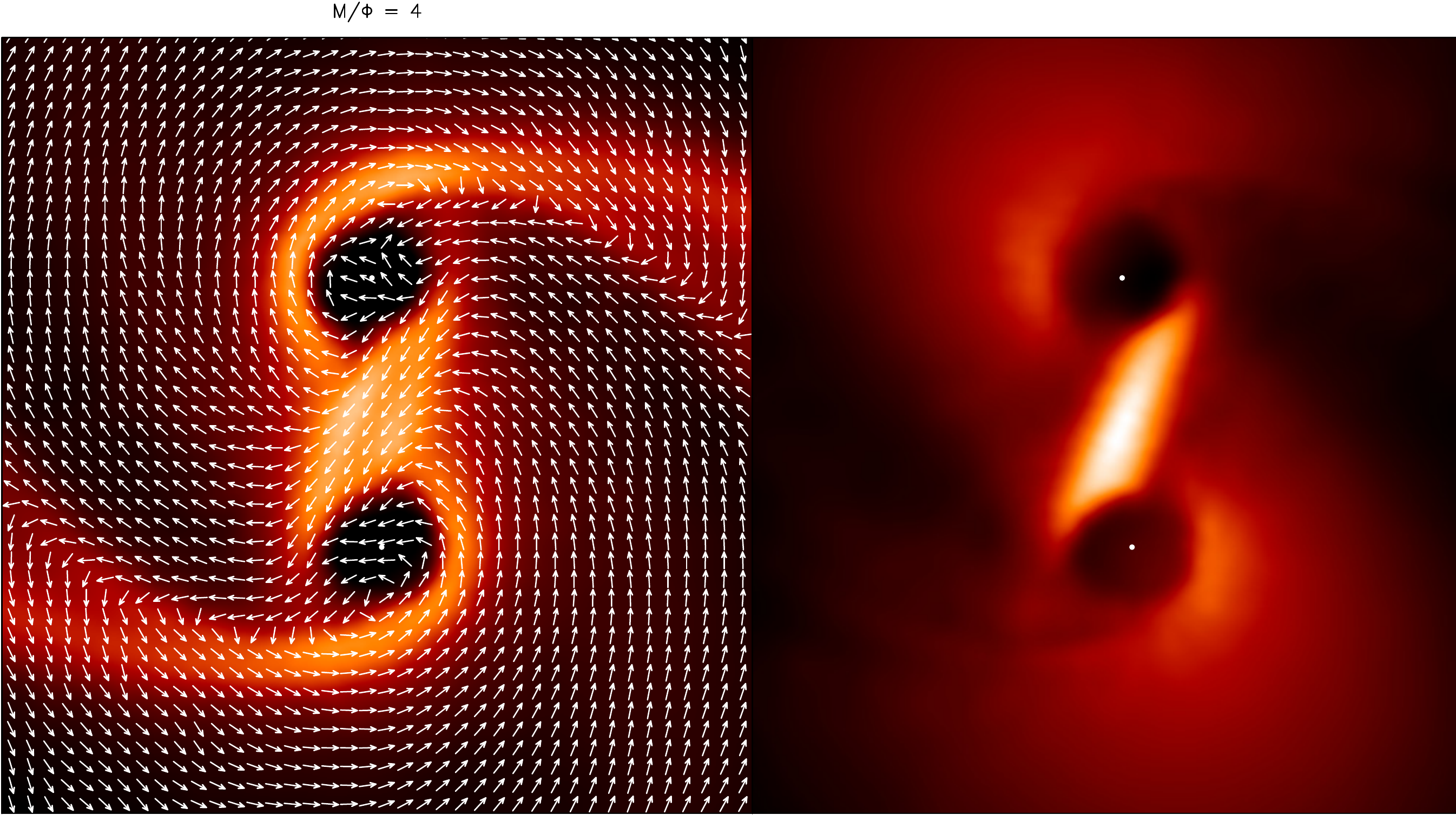}
\caption{\label{fig:cushion} Two panels show magnetic cushioning in the $M/\Phi = 4$ run. This example shows the system at $t = 1.35 t_{\text{ff}}$, thus corresponding to panel 5 in Fig. (\ref{fig:mtof4}). The panel on the left-hand side shows column density and integrated magnetic field vectors which, for better visibility, were set to equal magnitude. The right-hand side shows the integrated magnetic pressure. It can be seen, that the magnetic field forms a cushion, which prevents the two objects from merging. This figure might be compared to the similar Fig. 11 in Price \& Bate (2007), which shows the same quantities.}
\end{center}
\end{figure*}

\begin{figure*}
\begin{center}
\includegraphics[width=\textwidth]{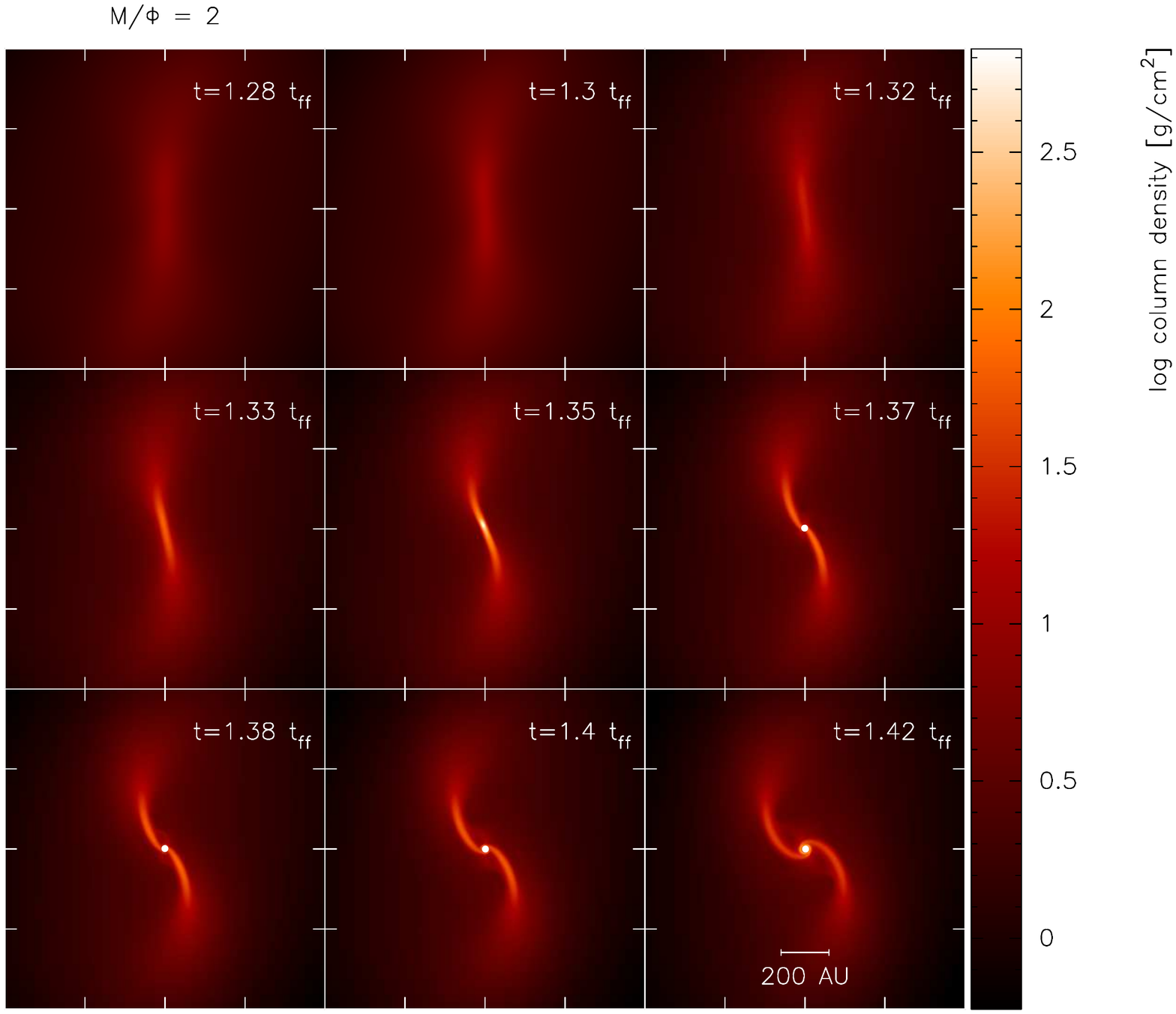}
\caption{\label{fig:mtof2} Simulation results with magnetic field parallel to the rotation ($z$) axis, with an initial field strength of $B_0 = 407\,\umu\unit{G}$. This corresponds to a mass-to-flux ratio of $M/\Phi = 2$, in multiples of the critical value. As previously, in horizontal direction, $x$-coordinates are plotted, and $y$-coordinates are plotted vertically. Shown is the column density, here integrated along the $z$-axis, in physical coordinates. The panels display a sequence of increasing time, measured in in units of free-fall times, $t_{\text{ff}}=2.4\times 10^{4}\,\unit{yr}$.} 
\end{center}
\end{figure*}

\section{INITIAL CONDITIONS}

For comparison with \cite{Price2007tg}, we chose the same initial setup as in their work. The initial cloud core has a spherical shape with a radius $R = 4 \times 10^{16} \unit{cm}$ and a mass $M=1\, \text{M}_{\odot}$, i. e.  we adopt a constant initial density $\rho_0 = 7.43 \times 10^{-18} \unit{g\,cm}^{-3}$ and a free-fall time

\begin{equation}
t_{\text{ff}} = \sqrt{\dfrac{3\pi}{32G\rho_0}} \simeq 2.4 \times 10^4\,\unit{yr}.
\end{equation}

The initial setup is realized by distributing the particles on a closed-packed lattice. This kind of particle distribution ensures very good settling properties with a very low initial scatter. In Fig. (\ref{fig:scatter}) we show the running average of the density, as obtained from an initial snapshot of the system. An analysis by fitting the points shows, that the deviation from the analytical density $\rho_0$ over the whole cloud is only $0.2 \%$. So we conclude, that the influence of Poission noise, which, as discussed in \cite{Cartwright2009kh}, imposes serious problems for SPH estimators, is effectively reduced in our initial conditions. Additionally, we would like to point out that the particle distribution is identical for every considered model, as are all other physical parameters not related to magnetic fields.

The number of particles in the cloud are $300\,914$. As pointed out in \cite{Price2007tg}, these are about ten times as many particles as required from the Jeans resolution criterion \citep{Bate1997sj} for the chosen equation of state. For the MHD calculations, this cloud was embedded into a uniform, low-density medium with a temperature 30 times higher than within the cloud, so that cloud and medium are initially in pressure equilibrium. This approach has the advantage, that the magnetic field lines behave regularly at the cloud boundaries and thus cloud particles are prevented from being ejected into space by magnetic forces, which otherwise would be induced by ill-defined behaviour of the magnetic field at the cloud surface. The ambient medium is represented by $146\,074$ particles which does not add significant extra computational cost. Note, however, that for larger systems typically studied in the context of star cluster formation, this approach is not efficient any more, and another strategy has to be employed by, e. g., removing particles far away from the region of  interest. To avoid dilution of the medium, the whole system is placed into a cubic box with periodic boundary conditions and a side length of four times the cloud radius.

We consider a variation of the 'standard isothermal test case' [see \cite{Boss1979fk}, but in an SPH context also \cite{Bate1997sj}] which, in the hydrodynamical case, is known to lead to formation of binary stars, or, dependent on the concrete realization, even to multiple systems \citep[e. g.][]{Arreaga-Garcia2007zv}. Here, the initial density is altered by a non-axisymmetric $m=2$ perturbation in density,

\begin{equation}
\rho = \rho_0 \left[ 1+A\cos(2\phi)\right].
\label{eq:denspert}
\end{equation}
where the $\phi$ is the azimuthal angle with respect to the rotation axis. Realizations of non-uniform density distributions for regular spaced particles, however, are not easy to achieve. In this case, one possible way would be the use of particles with unequal masses. But this can lead to undesirable side-effects \citep[e.\ g.][]{Rosswog2009qo}, so this approach is not used by many researchers, except when the system under consideration is strongly centrally condensed \citep[e.\ g.][]{Arreaga-Garcia2010zh}. 
For small perturbations,  it is sufficient to slightly perturb the initial positions of the particles, in order to match the desired density distribution. To do so, we consider the linearized continuity equation, which reads

\begin{equation}
\delta\rho + \rho_0 \bnabla \cdot \delta\rbold = 0.
\end{equation}
From this it follows, by considering spherical polar coordinates and performing a straightforward integration [using eq.\ (\ref{eq:denspert})], that the perturbation in the azimuthal angle is given by
\begin{equation}
\delta\phi  = -\dfrac{A\sin(2\phi_0)}{2}.
\end{equation}
For the calculations presented in this work, $A=0.1$ has been chosen to make comparisons to other work possible.

The cloud has a solid-body rotation with an angular velocity of $\Omega = 1.006 \times 10^{-12}\,\unit{rad}\,\unit{s}^{-1}$ and we fix the initial temperature to $T=8.4\,\unit{K}$ and the mean molecular weight to  $\umu_{\text{mol}} =2$. With these choices, we get a sound speed of $c_{\text{s}} = 0.19\, \unit{km}\,\unit{s}^{-1}$ and initial energy ratios

\begin{align}
\alpha_{\text{therm}} &= \dfrac{E_{\text{therm}}}{\left| E_{\text{grav}}\right|} = 0.26\nonumber\\
\beta_{\text{rot}} &= \dfrac{E_{\text{rot}}}{\left| E_{\text{grav}}\right|} = 0.16.
\end{align}
An established quantitative measure for magnetic field support in self-graviting fluids is the mass-to-flux ratio. For a spherical cloud, it is given by

\begin{equation}
 \left(\dfrac{M}{\Phi}\right) \equiv \dfrac{M}{\pi R^2 B_0}
\end{equation}
and the critical value for this geometry \citep[see][]{Mouschovias1976dz} in cgs units,
\begin{equation}
 \left(\dfrac{M}{\Phi}\right)_{\text{crit}} \simeq 0.125\, \sqrt{\dfrac{1}{G}} \simeq 486\,\unit{g}\,\unit{cm}^{-2}\,\unit{\umu G}^{-1}.
\end{equation}
Below this value, the magnetic field is able to support the cloud against gravity, while above it, gravity will dominate the magnetic field. At this point, we would like to stress the fact, that the mass-to-flux ratio is the main parameter that is changed in this work. Using the mass-to-flux ratio in units of the critical value, the magnetic field strength is given by

\begin{equation}
B_0  = 814\, \umu\unit{G}\, \left(\dfrac{M}{\Phi}\right)^{-1} \left(\dfrac{M}{\text{M}_{\odot}}\right) \left(\dfrac{R}{4 \times 10^{16} \,\unit{cm}}\right)^{-2},
\end{equation}
respectively. In our calculations, the initial magnetic field is aligned with the rotation (z) axis. 

\section{RESULTS}
We chose models with several initial mass-to-flux ratios also considered in \cite{Price2007tg}, summarized in Table (\ref{tab:magprops}), together with the corresponding field strengths $B_0$. At this point, we would like to emphasize, that the mass-to-flux ratio is given in units of the critical value throughout the subsequent sections. Furthermore, the table includes two additional columns containing the ratio of gas pressure to magnetic pressure,
\begin{equation}
\beta_{\text{plasma}} = \dfrac{P_{\text{gas}}}{B_0^2/(8\pi)}
\end{equation}
and the Alfv\'{e}n speeds
\begin{equation}
v_{\text{A}} = \dfrac{B_0}{\sqrt{4\pi\rho_0}},
\end{equation}
respectively.

\begin{table}
\caption{Initial parameters characterizing the magnetic field within the spherical cloud for each considered model.}
\label{tab:magprops}
\begin{tabular}{@{}rrrr}
\hline  $M/\Phi$ & $B_0\,[\umu\unit{G}]$ & $\beta_{\text{plasma}}$ & $v_{\text{A}}\,[\unit{km/s}]$\\ 
\hline\hline $\infty$ &  $0$ &$ \infty$ & 0 \\ 
$20$     & $40.7$ & $39$ &  $0.04$\\ 
$10$     & $81.3$ &  $9.80$ &  $0.08$\\ 
$7.5$    & $108.5$ & $5.50$  & $0.11$\\ 
$4$       & $203$ &  $1.57$&  $0.21$\\
$2$      &  $407$ & $0.39$ &  $0.42$\\
$1$      & $814$ &  $0.09$ & $0.84$\\
\hline
\end{tabular}
\medskip
\newline
$M/\Phi$ is the mass-to-flux ratio measured in critical units, $B_0$ is the initial magnetic field strength, $\beta_{\text{plasma}}$ is the ratio of the gas pressure to the magnetic pressure, and $v_{\text{A}}$ is the Alfv\'{e}n speed.
\end{table}

\begin{figure*}
\begin{center}
\includegraphics[scale=0.7]{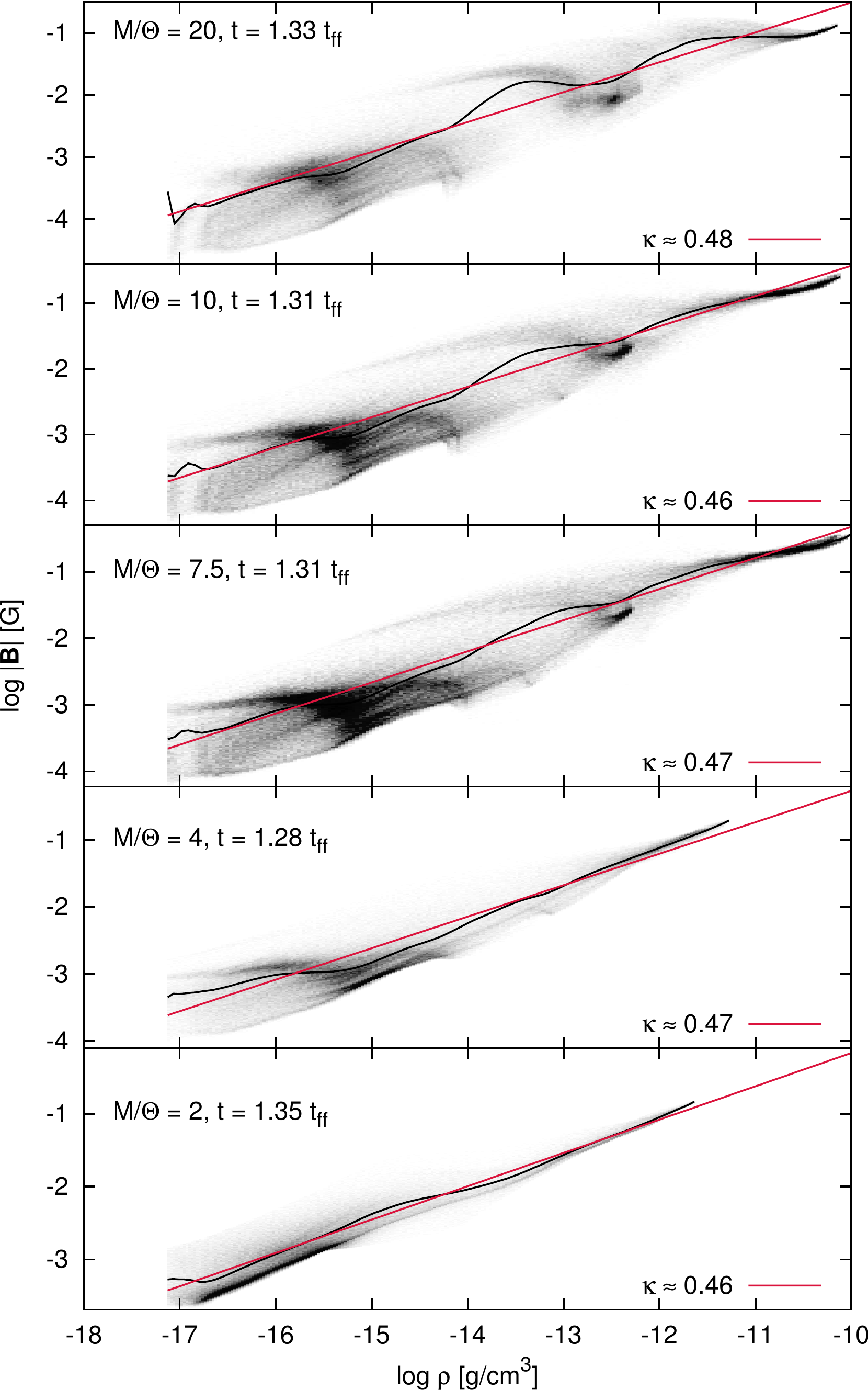}
\caption{\label{fig:brho} Panel of histogram plots showing the  $B-\rho$ relation for each initial mass-to-flux ratio where only the last snapshot prior to sink particle formation is considered. The grey-scaled 2D histogram indicates the value of magnetic field strength $B$ as function of $\rho$, where the colour intensity is proportional to the number of particle counts within the bin considered. The solid black line is the moving average calculated from the contributions by the particles. The red line shows a fit assuming the relation $B\propto \rho^{\kappa}$, where the value of $\kappa$ obtained by the fit is displayed on the lower right of each plot. Only particles with densities above $\rho_0$ were considered in these plots.} 
\end{center}
\end{figure*}

\subsection{Column density evolution}

Figure  (\ref{fig:hydro}), where the hydrodynamical case is shown, displays a sequence of nine plots showing the column density integrated along the $z$-direction, i. e. parallel to the rotation axis. The timing is very similar compared to \cite{Price2007tg}, with a delay in our calculations estimated to be below $5 \%$. Since such a disagreement in comparisons to other work is frequently reported in the literature, e. g. in \cite{Commercon2008fc}, we regard the difference here as being acceptable. Furthermore, we find that orientation and size of the spiral pattern agree quite well with those found in \cite{Price2007tg}. Especially their result shown for $t = 1.30 t_{\text{ff}}$ seems to coincide with our result at $t=1.35 t_{\text{ff}}$. Also we notice, that rotational symmetry is well preserved. From these observations we conclude, that our hydrodynamical simulation results are well in agreement with other results obtained by other authors using a similar setup. 

The case with $M/\Phi=20$, which corresponds to a initial field strength of $B_0 = 40.7\,\umu\unit{G}$, is shown in Figure (\ref{fig:mtof20}). This is a comparably weak field, and so there are almost no changes in the spiral patterns or in timing, compared to the pure hydrodynamical case. Also the onset of star formation is not significantly hindered. Only in the later stages, displayed in the lower three panels of Figure (\ref{fig:mtof20}), we notice a small speed-up in the dynamics.

The Figures (\ref{fig:mtof10}) and (\ref{fig:mtof7.5}), showing the case with $B_0 = 81.3\,\umu\unit{G}$ ($M/\Phi = 10$) and $B_0 = 108.5\,\umu\unit{G}$ ($M/\Phi = 7.5$), respectively, are more interesting. Here we see, that the collapse leads to the formation of protostars already in earlier stages of the simulation than in the cases considered before. We estimate the speed-up to be about $t=0.02\, t_{\text{ff}}$. Additionally, we see that at $t=1.38\, t_{\text{ff}}$ in the $M/\Phi = 10$ and at $t=1.37\, t_{\text{ff}}$ in the $M/\Phi = 7.5$ case, a third star is formed which in the further evolution turns out to be gravitationally bound to one of the other protostars, respectively. 

In order to explain the formation of a triple system in the latter two cases, we performed additional simulations in these cases. These simulations, however, do not include magnetic tension forces. That is, the $B^k B^l$ terms in the magnetic stress tensor, eq. (\ref{eq:tensor}), were neglected. In these cases (not shown) we find that binary systems are formed, but no triple system. Thus, magnetic tension could be a possible reason for this behaviour. However, fragmentation in the isothermal regime is extremely sensitive to small perturbations, so small differences in the fragmentation behaviour should be viewed with caution. It is also known that use of a barotropic equation of state can severely overestimate the amount of secondary fragmentation compared to more realistic calculations where radiative heating of the gas is explicitly accounted for \citep[e.\ g.][]{Offner2009fk,Price2009ys,Peters2010uq,Peters2010kx,Peters2010fk}.

More interesting, with respect to magnetic tension, is the case with mass-to-flux ratio of $M/\Phi = 4$, corresponding to $B_0 = 203\,\umu\unit{G}$, where we see a different picture (Figure \ref{fig:mtof4}). While star formation sets in still earlier than in the previous case, it is accompanied with a pronounced filamentary structure, already visible at $t=1.28\, t_{\text{ff}}$. Contrary to the two cases considered before, only a binary system is formed here, where \cite{Price2007tg} find just a single star in this case. This can be attributed to the 'magnetic cushioning' effect, shown in detail in Fig. (\ref{fig:cushion}) at $t=1.28\, t_{\text{ff}}$ thus corresponding to panel 5 in Figure (\ref{fig:mtof4}). This 'magnetic cushion', due to tension forces, prevents the two protostars from merging into a single one, what otherwise is likely to happen given the close encounter of the two objects. It must be noted, however, that \cite{Price2007tg} find this 'magnetic cushion' only in cases where the initial magnetic field was perpendicular  to the rotations axis. A possible reason for this is an underestimation of this effect in their calculations with the magnetic field parallel to the rotation axis, due to the restrictions of the Euler potentials in capturing certain field geometries. 

Figure (\ref{fig:mtof2}) shows the case with a very strong field, $B_0 = 407\,\umu\unit{G}$ ($M/\Phi = 2$). Here we see a bar structure forming, condensing to a single star forming its central region. This protostar, however, is formed comparably late compared to the cases with weaker field strengths. Finally, we note that in the critical case with $M/\Phi = 1$, no star is formed at all (not shown).

\subsection{Relationship between $B$ and $\rho$}
Observations \citep[e. g.][]{Crutcher1999ys,Heilesuq} as well as theoretical investigations and simulations \citep[e. g.][]{Mouschovias1976vn, MOUSCHOVIAS1991zr,Fiedler1993kx,Desch2001fk,Li2004uq} suggest a scaling behaviour of $B$ with the density $\rho$, which is usually parametrized as $B\propto\rho^{\kappa}$.

For an isothermal (or sufficiently cooled) core, this relation with $\kappa = 1/2$ can be motivated 
by assuming, that magnetic fields can not provide support against gravity along the field lines (i. e. parallel to the $z$ axis), leading to a disc like morphology of the cloud in later stages of the collapse. By further taking the validity of flux-freezing in ideal MHD into account, one yields $B\propto\sqrt{\rho T}$ \citep[e. g.][]{Heilesuq}, reducing to $B\propto\sqrt{\rho}$ in the isothermal case.

On the other hand, for gravity exceeding both magnetic and turbulent support, as well as for negligible angular velocity, the cloud could also collapse more rapidly \citep[e. g.][]{Heilesuq}. For the case of a weak field and a spherical cloud, \cite{Mestel1956kx} performed an analysis showing that the morphology is almost unaffected in such a case; furthermore they obtained a value of $\kappa=2/3$. 

In order to examine which one of these cases is realized in our simulation results, we show in Fig. (\ref{fig:brho}) the dependence of the magnetic field strength $B$ on the cloud density $\rho$ for each of the considered cases at a time late in the evolution of the cloud, before the first sink particle is created. The grey-scaled two-dimensional histogram shows the magnetic field strength, with the colour intensity proportional to the number of particles within each bin. For the calculation of each histogram, $200 \times 200$ bins were considered, respectively. The black solid curve indicates the moving average while the red solid curve was obtained from a fit.  Our fit shows a power-law behaviour with a value for $\kappa$ which is, in each case, below but close to a value of $1/2$ indicating the emergence of a disc like geometry during the collapse, in agreement with other studies \citep{Fiedler1993kx,Desch2001fk,Li2004uq}.  But note, that the works by \cite{Banerjee2006uq} and \cite{Price2007tg} report values of $\kappa \approx 0.6$ more closely to a value of $\kappa=2/3$, thus indicating a more spherical collapse. However, our initial angular velocity $\Omega$ is rather high compared to the latter works, who analysed the $B-\rho$ relation only in an unperturbed setup with angular velocities of $1.89 \times 10^{-13}$ \citep{Banerjee2006uq} and $1.77 \times 10^{-13}\,\unit{rad}\,\unit{s}^{-1}$ \citep{Price2007tg}, respectively. So in our case, flattening of the cloud during the collapse is expected to be enhanced compared to the latter works, especially since the effects of magnetic braking, see next sub-section, are found to be rather weak. 
 
\subsection{Angular momentum transport}
\begin{figure}
\begin{flushleft}
\includegraphics[width=9cm]{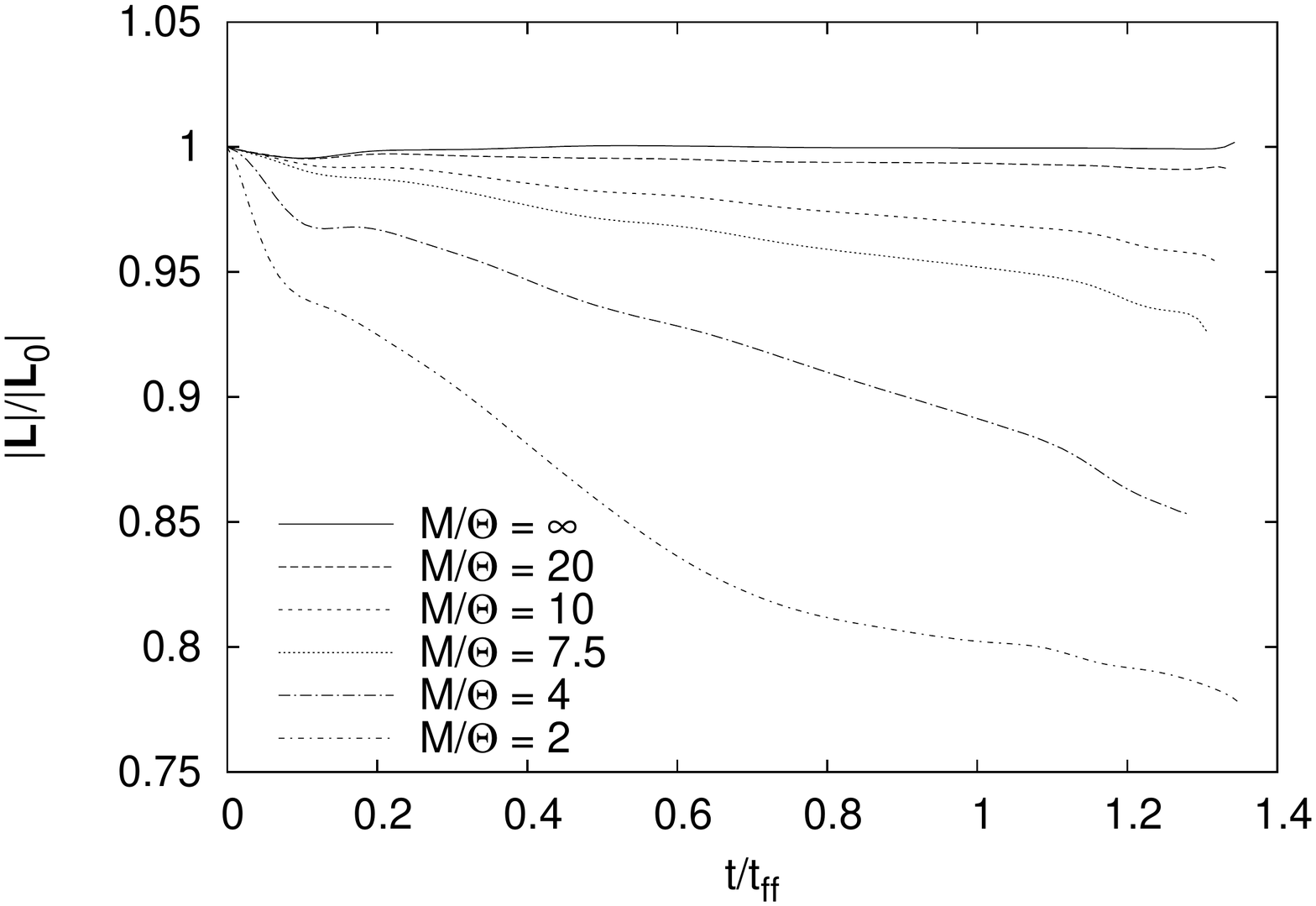}
\caption{\label{fig:angmom} Angular momentum evolution for each cloud, distinguished by its initial mass-to-flux ratio, respectively. Shown is the time evolution (in $t_{\text{ff}}$) of the normalized angular momentum, $|\mathbf{L}|/|\mathbf{L}_0|$, measured within the initial cloud radius $R$. The plot quantifies the decay of the initial angular momentum, dependent on the initial mass-to-flux ratio.}
\end{flushleft}
\end{figure}

Also of interest is the influence of magnetic fields on the rotation of the cloud, usually attributed to a process called magnetic braking. The usual qualitative picture describing this process is, that torsional Alfv\'{e}n waves are launched into the ambient medium, if the latter has a different rotation than the cloud. Thus, the cloud is slowed down by transport of angular momentum outwards \citep[e. g.][]{Mouschovias1979zr,Mouschovias1980ly,Mestel1984ve,MOUSCHOVIAS1991zr}. Since the ambient medium  has no initial velocity at all, it can be expected that magnetic braking takes place in our models and shows measurable effects within the simulation time. To quantify the effect of magnetic braking, we follow the evolution of the normalized angular momentum, $|\mathbf{L}|/|\mathbf{L}_0|$, within the initial cloud radius $R$, shown in Fig. (\ref{fig:angmom}). 

For a more quantitative analysis, we consider the timescale characteristic to magnetic braking, $\tau_b$, which is the time the outward propagating Alfv\'{e}n waves need to enfold a fraction of the ambient gas corresponding to a moment of inertia equal to the cloud. In the case of a spherical cloud, threaded by a uniform magnetic field parallel to the rotation axis, the braking time can be estimated by \citep{McKee1993ab}

\begin{equation}
\tau_b = \dfrac{8}{15} \left( \dfrac{\rho_0}{\rho'} \right) \dfrac{R}{v'_{\text{A}}}
\end{equation}
where the primes denote values in the ambient medium. But note that the applicability of this classical analysis \citep[e. g.][]{MOUSCHOVIAS1991zr} was criticised by \cite{Hennebelle2009kx}, since the magnetic braking within a collapsing core might not be fully captured by this analysis. However, since we do not concentrate on the details of disc formation, but on the whole cloud, we expect that this analysis still gives a rough approximation of  timescales related to magnetic braking.  By inserting numerical values, we see that the braking times are distributed, monotonically decreasing with decreasing mass-to-flux ratio, in a range from $\tau_b \approx 35\,t_{\text{ff}}$ for  $M/\Phi=20$, to $\tau_b \approx 3.5\,t_{\text{ff}}$ for $M/\Phi=2$. This approximation seems to be in good agreement with our results as displayed in Fig. (\ref{fig:angmom}) which show a rather slow braking which can be expected for supercritical clouds with a high initial density ratio $\rho_0/\rho'$ \citep{McKee1993ab}. Furthermore, we would like to emphasize that our models are based on an ideal, one fluid MHD formulation which represents a fully ionized plasma, and therefore does not allow for ambipolar diffusion which limits the efficiency of magnetic braking considerably, as was pointed out already by \cite{Hosking2004kx}. Also of importance for the efficiency of magnetic braking is the initial field geometry. It was shown by \cite{Price2007tg}, that an initial magnetic field perpendicular to the rotation axis increases the efficiency of magnetic braking substantially which is attributable to magnetic tension. However, \cite{Hennebelle2009kx}, who investigated collapse problems with  magnetic fields inclined to the rotation axis systematically, propose that increasing the inclination angle of the magnetic field reduces the efficiency of magnetic braking.

\subsection{Numerical stability of the SPMHD algorithms}

\begin{figure*}
\begin{center}
\includegraphics[scale=0.7]{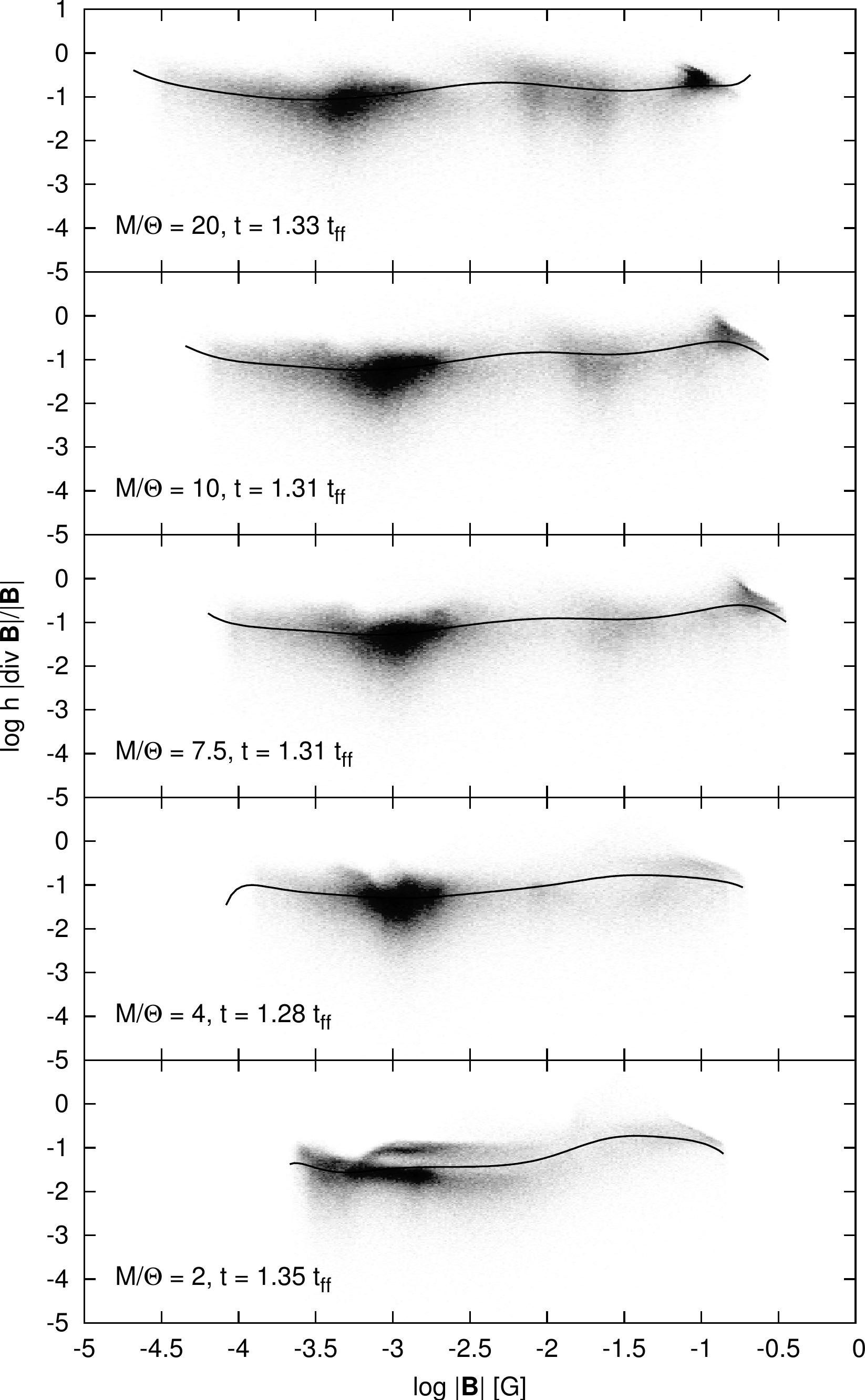}
\caption{\label{fig:divb} Numerical divergence plotted as function of the magnetic field strength. The grey-scaled 2D histogram shows the values of $h |\bnabla\cdot\Bbold|/|\Bbold|$ as a function of the magnetic field for particles with a density larger than $\rho_0$. The black solid line represents the moving average of the numerical divergence obtained from the individual values of the particles. As in Fig (\ref{fig:brho}), the last snapshot prior to sink particle formation is considered.} 
\end{center}
\end{figure*}

A further aspect of our work considers the influence of divergence of the magnetic field on our results. Since non-vanishing divergence of the magnetic field used to impose serious constraints on the usability of MHD in SPH, in particular in a star formation context, we are going to discuss the implications of this issue in some detail. Especially, since the meaning of the divergence is probably mistaken, we are going to clarify matters based on a similar chain of arguments as given in \cite{Kotarba2010ab}.

The SPH estimator for the divergence of the magnetic field at the position of particle $i$ given by, e. g., 
\begin{equation}
\left( \bnabla \cdot \Bbold \right)_i = -\dfrac{1}{\rho_i} \sum_j m_j \left( \Bbold_i -\Bbold_j \right) \cdot \bnabla_i W_{ij}
\end{equation}
calculates the weighted contribution of the differences of the magnetic field due to the $N$ neighbouring particles to particle $i$ within a smoothing length $h$. So the magnitude of the divergence calculated this way, is essentially based on the (irregular) distribution of the particles and thus a measure of sub-smoothing-length fluctuations of the field. It must be noted, however, that this \emph{numerical divergence}, which is not a \emph{physical divergence} caused by magnetic monopoles, is also present in Euler potentials. The latter are free of physical divergence by definition, but show a mean numerical divergence, measured by the expression $\left< h|\bnabla \cdot \Bbold| \right> / \left< |\Bbold| \right>$, that can approach values of order unity during a simulation as reported in \cite{Kotarba2009fk,Kotarba2010ab}. Additionally, the correction techniques employed within the present SPMHD scheme ensure, that the magnetic field evolution is not affected directly by the measured numerical divergence. However, it is of course advisable to keep it as low as possible, and to keep track of the value of numerical divergence within simulation time, to ensure that irregularities in the results are not correlated with high values of numerical divergence.

To quantify the analysis of the effects of numerical divergence, we plotted it in Fig. (\ref{fig:divb}) as function of $|\Bbold|$ in a 2D histogram, where the intensity is proportional to the number of particles within a bin, with $200 \times 200$ bins in total. In this plot, the last snapshot before sink formation is considered, respectively, and only particles with densities larger than $\rho_0$ are taken into account. It can be seen, that the numerical divergence is distributed on a wide range of values, as in a similar analysis carried out by \cite{Kotarba2010ab} for their systems. Additionally, no strong dependence of the (mean) divergence on the strength of the magnetic field, and thus on the density, is visible. So since the curve of the mean divergence has a trend which is qualitatively the same for all initial mass-to-flux ratios, but the actual physical behaviour in the evolution of each setup shows huge differences, as illustrated above, we conclude that our results are meaningful and are not correlated to the value of numerical divergence.

\section{DISCUSSION}
We have performed a study on the influence of magnetic fields on collapse and fragmentation of a rotating molecular cloud core with an initial $m=2$ density perturbation and the initial magnetic field aligned with the rotation axis. The amplitude in each case was chosen to be 10 per cent, as commonly used in the literature. 

Since our approach is based on a induction equation formulation of SPMHD, in contrast to the Euler equations based approach used by \cite{Price2007tg} for their star formation calculations, we would like to emphasize that it is not a priori clear that our approach should work at all for collapse problems, given the considerable amount of failed attempts in using this method (Daniel Price, private communication). Most of those attempts showed a disruptive behaviour at large densities, attributed to high values of numerical divergence of the magnetic field. However, since the details of those simulations are not known to us, we can only speculate what the reasons for these differences might be. First of all, the used parameters controlling artificial viscosity and resistivity, respectively, have values which are not uncommon in the literature \citep[e. g.][]{Price2005ve,Dolag2009uq}, and thus are unlikely to have dramatic effects on the global evolution of our simulations. We also investigated the influence of replacing high-density regions by accreting sink particles. Therefore, we performed simulations without any sink particles and thus followed the evolution of our systems as long as the global timestep allowed this at reasonable computational cost, and we have not recognized any signs of disruptive behaviour there. However, since we do not know of other work that has used the regularization method by \cite{Borve2001zt} for collapse simulations before, we suppose that this method, in combination with artificial resistivity, could be more effective than other methods in preventing numerical divergence from corrupting the magnetic field evolution. 

Considering the column density evolution in our models, our results suggest an overall agreement with the findings from \cite{Price2007tg}. For very weak field strengths, as expected, there are almost no deviations from the pure hydrodynamical case. On the other hand, for a very strong field with $M/\Phi = 2$ star formation is substantially delayed and just a single star is formed.
 However, for intermediate field strengths with $M/\Phi = 10$, $7.5$, we actually see the formation of a triple system which is not present in the work by \cite{Price2007tg}, at least at these early stages. Using simulation without magnetic tension, the third protostar did not form but most probably due to small perturbations, to which a barotropic equations of state is very sensitive, changing the sub-fragmentation pattern in the considered systems. So we conclude, that these differences in sub-fragmentation are probably not very meaningful. A larger difference between our work and \cite{Price2007tg} can be seen in the case of $M/\Phi = 4$, where a single star is formed in their calculations, but a binary system in our case. This difference is due to the 'magnetic cushioning effect' which is probably underestimated in their calculations with initial magnetic field parallel to the rotation axis, due to the intrinsical limitations of the Euler potentials. 

However, it is also instructive to compare our results to other findings in the literature. \cite{Hosking2004kx} investigated collapse and fragmentation using a two-fluid model, allowing them to model effects from non-ideal MHD. However, they started from sub-critical cores that became critical during the evolution via ambipolar diffusion. So it must be noted, that their investigations are quite different from our approach. They found no fragmentation in their magnetized models, but due to the different approaches used, it is difficult to relate their findings to our results. 

Furthermore, we would like to mention the work by \cite{Machida2005cd}, who investigated a large range of parameters in their fragmentation problems. However, since they started out from a filament with more complex initial perturbations in the density as well as in the magnetic field itself, their initial conditions are very different from those used in this work. So any comparison can be only of qualitative nature. Their characterizing parameter $\omega$ corresponds to $\sqrt{\beta_{\text{rot}}}=0.4$, while their $\alpha$ is equal to $v_{\text{A}}^2/c_{\text{s}}^2$. Thus, our models in a range from $M/\Phi=20$ to $M/\Phi=4$ are located in the 'vertical collapse region' in their Figure 10, were fragmentation is possible according to their analysis. The very strong field models with $M/\Phi=2$ to $M/\Phi=1$ are outside the horizontal range of their Figure 10, but the rather extreme values of $\alpha=4.88$ and $\alpha=19.54$, respectively, lead us to the speculation that they would be located in the region were no fragmentation occurs. Therefore, we find that our results globally agree with those found by \cite{Machida2005cd}.

 A comparison to \cite{Hennebelle2008fk} is quite difficult, since they use a different barotropic equation of state with a critical density of $\rho_{\text{crit}} = 10^{-13}\unit{g}\,\unit{cm}^{-3}$, the latter being one order of magnitude higher than ours. Additionally, their $\beta_{\text{rot}}$ has a value of $0.045$ lower than in our models. In their weak perturbation models using $A=0.1$, they find no fragmentation for $M/\Phi \leq 20$, thus all of their models form a single star. This is, with the exception of the $M/\Phi=2$ case, in disagreement with our findings. However, this not surprising because of the differences to their initial setup.
 
\cite{Ziegler2005fk} and \cite{Fromang2006kx}, investigated the case $M/\Phi=2$ using the same initial conditions, which are very similar to those used by \cite{Hennebelle2008fk}. They also used the same equation of state as \cite{Hennebelle2008fk} and $\beta_{\text{rot}}=0.045$, thus only a qualitative comparison is possible to our work. \cite{Ziegler2005fk} finds formation of a binary in this case, while \cite{Fromang2006kx} get different results depending on the flux solver used, namely no binary with the Lax-Friedrich solver and a binary with the Roe solver. However, the latter binary merged to a single fragment shortly thereafter. Our results for $M/\Phi=2$ show no sign of binary formation, so considering this particular case our results show more similarities with \cite{Fromang2006kx} than with \cite{Ziegler2005fk}. 

Furthermore, we also investigated for each mass-to-flux ration the dependence of the magnetic field strength $B$ on the density $\rho$ within the final stages of collapse within a core. Our results, showing a value of $\kappa$ close to $1/2$ in the power-law relation $B \propto \rho^{\kappa}$, are well in agreement with a picture with vanishing magnetic support parallel to the symmetry axis. Thus, the cloud finally ends in a disc-like morphology, independent of the initial mass-to-flux ratio. Such a behaviour is also frequently reported in the literature \citep{Mouschovias1976vn, MOUSCHOVIAS1991zr,Fiedler1993kx,Desch2001fk,Li2004uq}.

An additional investigation concerned the angular momentum transport, yielding that magnetic braking is weak in the models we considered, as could be expected from analytical reasoning \citep{Mouschovias1979zr,Mouschovias1980ly,Mestel1984ve,MOUSCHOVIAS1991zr,McKee1993ab}, and from non-ideal MHD simulations carried out by \cite{Hosking2004kx}. The influence of the initial field geometry on the efficiency of magnetic braking is currently under discussion. \cite{Price2007tg} advocate an increased efficiency with an an initial field perpendicular to the rotation axis, but \cite{Hennebelle2009kx} propose the opposite. However, we would like to emphasize, that we consider the whole cloud in this analysis and pay no attention on the impact of magnetic braking on disc formation. Thus we regard the recent criticism of the classical analysis by \cite{Hennebelle2009kx} as not influential to our analysis.

Finally, we would like to stress the fact, that according to our analysis of numerical divergence, we do expect that our results are not corrupted by artefacts and that therefore our results show the correct physical behaviour within our systems.

\section{SUMMARY}
In this work, we carried out magnetohydrodynamical computer simulations of the collapse of molecular cloud cores, initially disturbed with $m=2$ density perturbations.

The method is based on a formulation of smoothed particle magnetohydrodynamics (SPMHD) that evolves the magnetic field directly via the induction equation and thus does not utilize any form of scalar or vector potentials. Stability and noise reduction are ensured by techniques implemented by \citet{Dolag2009uq}, which have, to the best of our knowledge, not yet been applied in this combination to star formation problems.

From the results of this work, we draw several main conclusions. First, we find that our formulation of SPMHD did well in reproducing essential features obtained in other work with similar initial conditions, but using different methods. Thus we conclude that our approach is a viable scheme to attack star formation problems. Second, our results show good global agreement with the work by \cite{Price2007tg}, with the exception of cases with higher field strength where magnetic tensions aids binary fragmentation via the 'magnetic cushioning effect' in our simulations. This effect is not present in the corresponding results in \cite{Price2007tg}, most probably due to limitations of the Euler potentials approach in representing certain geometries of the magnetic field.

\section*{ACKNOWLEDGEMENTS}
 F.B.\ thanks Daniel Price for the routine which generates particles in a close-packed arrangement and many helpful discussions. Rendered plots were made using the {\small SPLASH} software written by Daniel Price \citep[see][]{Price2007hc}, available at \url{http://users.monash.edu.au/~dprice/splash}. Granting of computer time from John von Neumann-Institute for Computing (NIC), Jülich, Germany, is gratefully acknowledged.
 
  K.D.\ acknowledges the support by the DFG Priority Programme 1177 and additional support by the DFG Cluster of Excellence 'Origin and Structure of the Universe'.
   
 R.S.K.\ acknowledges financial support from the {\em Landesstiftung Baden-W{\"u}rttemberg} via their program International Collaboration II (grant P-LS-SPII/18) and from the German {\em Bundesministerium f\"{u}r Bildung und Forschung} via the ASTRONET project STAR FORMAT (grant 05A09VHA). R.S.K.\ furthermore gives thanks for subsidies from the {\em Deutsche Forschungsgemeinschaft} (DFG) under grants no.\ KL 1358/1, KL 1358/4, KL 1359/5, KL 1358/10, and KL 1358/11, as well as from a Frontier grant of Heidelberg University sponsored by the German Excellence Initiative. R.S.K. also thanks the KIPAC at Stanford University and the Department of Astronomy and Astrophysics at the University of California at Santa Cruz for their warm hospitality during a sabbatical stay in spring 2010. The KIPAC is sponsored in part by the U.S. Department of Energy contract no. DE-AC-02-76SF00515.
 
Finally, we would like to thank our referee, Daniel Price, whose remarks led to a significant improvement of this paper.
\bibliography{buerzleetal}
\label{lastpage}
\end{document}